\documentclass[amsmath,amssymb,twocolumn]{revtex4}
\usepackage{graphicx}
\usepackage{amscd,amsmath,amsthm,amsfonts,amssymb}

\def\ri{{\rm i}}

\def\re{{\rm e}}
\def\PT{$\mathcal{PT}$}

\def\[{\begin{equation}}
\def\]{\end{equation}}

\begin{document}
\title{Symmetry breaking with opposite stability between bifurcated asymmetric solitons in parity-time-symmetric potentials}
\author{Jianke Yang}
\affiliation{Department of Mathematics and Statistics, University of Vermont, Burlington, VT 05405, USA}




\begin{abstract}
We report a new type of symmetry-breaking bifurcation of solitons in optical systems with parity-time-symmetric potentials. In this bifurcation, the two bifurcated branches of asymmetric solitons exhibit opposite stability, which contrasts all previous symmetry-breaking bifurcations in conservative and non-conservative systems. We show that this novel symmetry-breaking bifurcation can be exploited to achieve unidirectional propagation of high-intensity light beams in parity-time-symmetric potentials.
\end{abstract}

\maketitle

Symmetry breaking of solitons is a common phenomenon in nonlinear wave systems \cite{Malomed_book}. In this symmetry breaking, two branches of asymmetric solitons emerge from the base branch of symmetric solitons through a pitchfork bifurcation, when the power of the symmetric soliton crosses a certain threshold. This bifurcation has been heavily studied, especially in the nonlinear Schr\"odinger (NLS)-type models with symmetric conservative potentials or symmetric gain-loss landscape \cite{Panos2005,Kartashov2011,Kirr2011,Yang2013}. It has been experimentally observed as well \cite{Panos2005}. In parity-time (\PT) symmetric complex potentials, while this bifurcation is less common, it has been reported in certain classes of \PT-symmetric potentials \cite{Yang2014}.

In all previous symmetry-breaking bifurcations, the two bifurcated branches of asymmetric solitons always featured the same stability. In systems with symmetric conservative potentials or symmetric gain-loss landscape \cite{Panos2005,Kartashov2011,Kirr2011,Yang2013}, this same stability between the two asymmetric solitons is guaranteed, because the system is symmetric in space and thus does not distinguish between the left and right. In \PT-symmetric potentials, while the system does distinguish between the left and right due to the anti-symmetric gain-loss landscape, the reported symmetry breaking in a Kerr nonlinear medium \cite{Yang2014} still exhibited the same stability for the two branches of asymmetric solitons.

In this article, we report a new type of symmetry breaking of solitons, where the two bifurcated branches of asymmetric solitons exhibit opposite stability. This symmetry breaking is discovered in an optical system where a light beam propagates in certain \PT-symmetric potentials under cubic-quintic or saturable nonlinearity.
This opposite stability indicates that the system favors one asymmetric soliton over the other. As an application of this bifurcation, we demonstrate a device which features unidirectional propagation for high-intensity beams but reciprocal propagation for low-intensity beams.

The optical system we consider is nonlinear paraxial beam propagation in complex \PT-symmetric potentials under cubic-quintic or saturable nonlinearity. We first consider cubic-quintic nonlinearity, where the mathematical model is
\cite{Kivshar_book,Yang_book}
\begin{equation} \label{Eq:NLS}
\ri \Psi_z + \Psi_{xx} + V(x)\Psi + |\Psi|^2 \Psi +\gamma |\Psi|^4\Psi = 0.
\end{equation}
Here, $\Psi$ is the complex envelope of the light's electric field, $z$ is the propagation distance, $x$ is the transverse
coordinate, $V(x)$ is a \PT-symmetric potential, i.e., $V^*(x)=V(-x)$, with the asterisk `*' representing complex conjugation,
and $\gamma$ is the coefficient of quintic nonlinearity. All variables have been non-dimensionalized. Physically, the \PT-symmetric potential means that the refractive index of the medium [i.e., Re($V$)] is symmetric in space, while the gain-loss profile [i.e., Im($V$)] is anti-symmetric in space \cite{Musslimani2008}. Such potentials have been fabricated in various physical experiments \cite{Segev2010,Xiao2016,PT_review,PT_book}. In addition, the cubic-quintic nonlinearity is common in many materials too.

Solitons in Eq. (\ref{Eq:NLS}) are sought of the form
\[
\Psi(x, z)=\psi(x)e^{\ri\mu z},
\]
where $\mu$ is a real propagation constant, and $\psi(x)$ is a
localized function which satisfies the equation
\[  \label{e:psi}
\psi_{xx}+V(x)\psi+|\psi|^2\psi+\gamma |\psi|^4\psi=\mu \psi.
\]
Since the potential $V(x)$ is \PT-symmetric, solitons in Eq. (\ref{Eq:NLS}) are generally also \PT-symmetric, i.e., $\psi^*(x)=\psi(-x)$, and these solitons form continuous families parameterized by the propagation constant $\mu$ \cite{PT_review,PT_book}. Symmetry-breaking of solitons is forbidden in generic \PT-symmetric potentials \cite{YangSAPM}. However, in a special class of \PT-symmetric potentials
\[ \label{Vx}
V(x)=g^2(x)+\alpha g(x) +\ri g'(x),
\]
where $g(x)$ is a real and even function and $\alpha$ a real constant, symmetry breaking of solitons is possible, where two branches of non-\PT-symmetric (asymmetric) solitons bifurcate out from the base branch of \PT-symmetric solitons when the power of the \PT-symmetric soliton crosses a certain threshold. Since $g(x)$ can be an arbitrary even function and $\alpha$ an arbitrary constant, this special class of potentials can still accommodate a wide range of refractive-index and gain-loss profiles. For Kerr nonlinearity ($\gamma=0$), symmetry breaking has been reported in \cite{Yang2014}. However, for the cubic-quintic nonlinearity ($\gamma\ne 0$), we will show that this symmetry breaking exhibits behaviors which have never been seen before.

To be concrete, let us take $\gamma=-0.1$, $\alpha=-0.5$, and
\[ \label{gx}
g(x)=A\left(e^{-(x-x_0)^2}+e^{-(x+x_0)^2}\right),
\]
with $A=2$ and $x_0=1.2$. The corresponding \PT-symmetric potential is displayed in Fig. 1(a). The real part of this potential (i.e., the refractive index of the medium) is double-humped and symmetric, and these two humps can be viewed as two waveguide channels, which are centered at $x=-x_0$ and $x=x_0$ respectively. The imaginary part of this potential (i.e., gain-loss profile of the medium) is anti-symmetric. Notice that the gain and loss is quite strong, but this potential is still below phase transition, i.e., its spectrum is all-real \cite{Musslimani2008,Bender1998}.

This \PT-symmetric potential admits a family of \PT-symmetric solitons, which bifurcates out from the linear discrete eigenvalue $\mu_0\approx 1.23$ of the potential. The power curve of this soliton family is shown as the solid line in Fig. 1(b), where the power is defined as $P(\mu)=\int_{-\infty}^\infty |\psi(x; \mu)|^2 dx$. At a point marked by letter `d' of this power curve, the corresponding soliton is shown in Fig. 1(d). It is seen that the amplitude function $|\psi(x)|$ of this soliton is symmetric since $\psi(x)$ is \PT-symmetric.

On this \PT-symmetric soliton branch, symmetry breaking occurs at $\mu_c\approx 1.65$, where two branches of asymmetric solitons bifurcate out on the right side of the bifurcation point. These two asymmetric solitons are related by $\psi_{1}^*(x, \mu)=\psi_{2}(-x, \mu)$. Thus, their power curves are identical and displayed as a dashed line in Fig. 1(b). At a point marked by letters `e,f' of this power curve, the corresponding two asymmetric solitons are shown in Fig. 1(e,f). The soliton in Fig. 1(e) has its energy concentrated in the right waveguide channel [i.e., the right hump of Re($V$)], while the one in Fig. 1(f) has its energy concentrated in the left waveguide channel. Obviously, neither soliton is \PT-symmetric.

The power curves in Fig. 1(b) cannot distinguish between the two branches of asymmetric solitons. Thus, in Fig. 1(c) we plot the soliton amplitudes $|\psi|$ at the right waveguide channel $x=x_0=1.2$ versus the propagation constant $\mu$. In this figure, the two asymmetric soliton branches (dashed lines), with energy concentrated in the right and left waveguide channels, are above and below the base \PT-symmetric soliton branch (solid line) respectively; thus they are now clearly distinguishable.

\begin{figure}[htbp]
\centering
\fbox{\includegraphics[width=\linewidth]{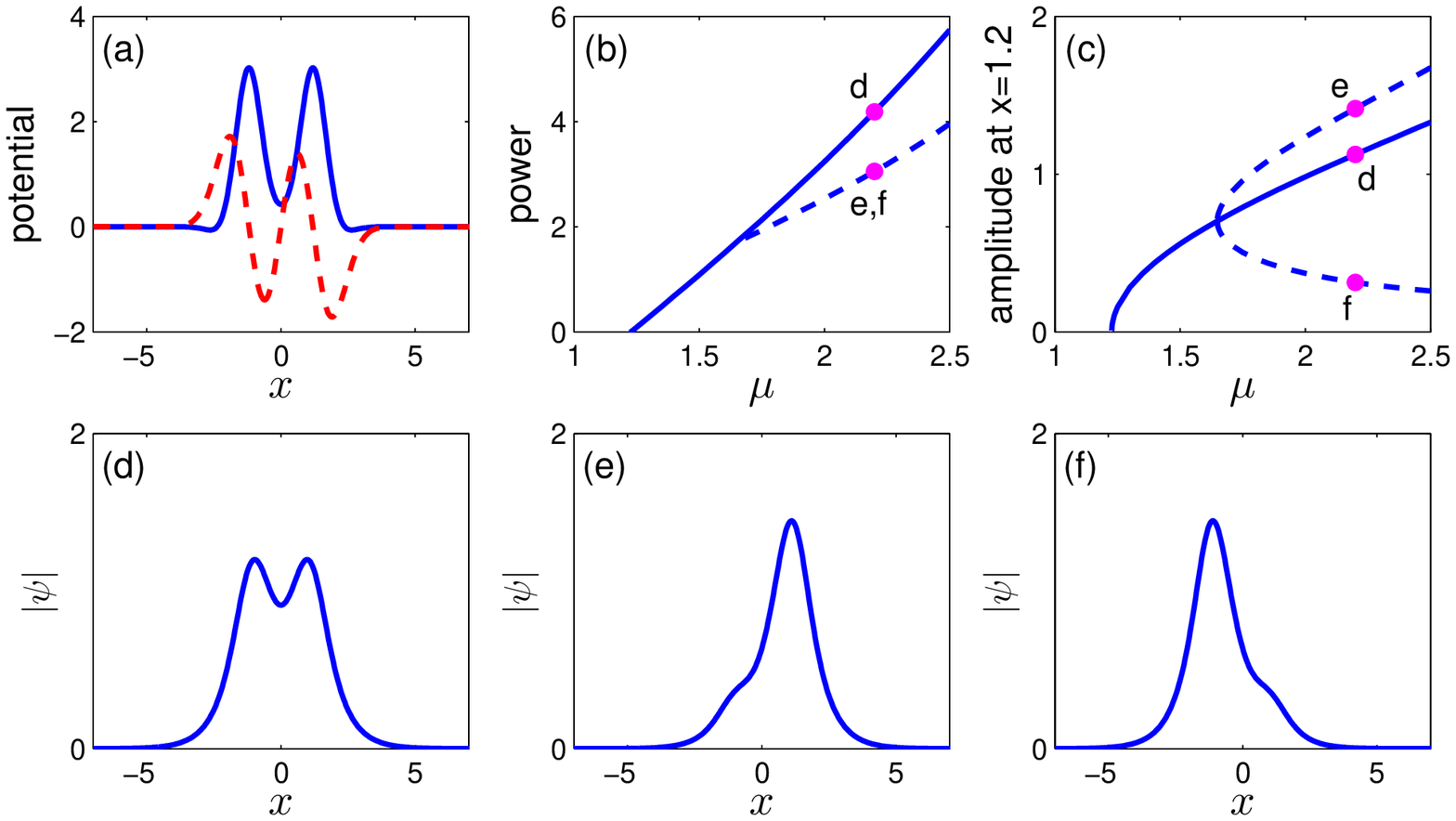}}
\caption{\PT-symmetry-breaking of solitons in the cubic-quintic model (\ref{Eq:NLS}) with $\gamma=-0.1$. (a) \PT-symmetric potential $V(x)$ given by equations (\ref{Vx})-(\ref{gx}) with $\alpha=-0.5$; solid blue: Re($V$); dashed red: Im($V$). (b) Power curves of \PT-symmetric (solid) and asymmetric (dashed) solitons in this potential. (c) The amplitude $|\psi|$ of the soliton at the right waveguide channel $x=x_0=1.2$ versus the propagation constant $\mu$. Solid: \PT-symmetric branch; dashed: asymmetric branches. (d,e,f) Profiles of soliton amplitudes $|\psi(x)|$ at parameter points marked by letters `d,e,f' in panels (b,c) respectively.}
\end{figure}

Our focus in this article is the stability behaviors of solitons near the symmetry-breaking point. In particular, we consider their linear stability. For this purpose, we perturb these solitons as
\begin{equation}
\Psi(x,z) = \re^{\ri \mu z} \left[ \psi(x) + u(x) \hspace{0.05cm}
\re^{ \lambda z} + w^*(x) \hspace{0.05cm} \re^{ \lambda^* z}
\right],
\end{equation}
where $|u|,  |w|\ll |\psi|$. Substituting it into Eq.~(\ref{Eq:NLS}) and linearizing, we arrive at the eigenvalue
problem
\begin{equation}
{\cal L} \left[\begin{array}{c} u\\ w
\end{array} \right] = \lambda   \left[\begin{array}{c} u\\ w
\end{array}\right],  \label{EigenProblem}
\end{equation}
where
\begin{align*}
{\cal L} &= \left[ \begin{array}{c c} L_{11} &  L_{12} \\
L_{12}^* & L_{11}^*\end{array} \right], \\
L_{11} &= \ri\left(\partial_{xx}  +  V -\mu + 2 |\psi|^2+3\gamma |\psi|^4\right),  \\
L_{12} &= \ri \psi^2(1+2\gamma |\psi|^2).
\end{align*}
If eigenvalues $\lambda$ with positive real parts exist, the soliton is linearly unstable; otherwise it is linearly stable.

It is easy to see that eigenvalues of operator ${\cal L}$ always come in conjugate pairs $(\lambda, \lambda^*)$. In addition, if two asymmetric solitons are related as $\psi_{1}^*(x, \mu)=\psi_{2}(-x, \mu)$, as those in Fig. 1, then eigenvalues for $\psi_2(x,\mu)$ would be negative of those for $\psi_1(x,\mu)$.

The eigenvalue problem (\ref{EigenProblem}) can be numerically computed by the Fourier collocation method \cite{Yang_book}. Our computation shows that, the \PT-symmetric branch of solitons in Fig. 1(b,c) is stable before the symmetry-breaking bifurcation, and becomes unstable after the bifurcation. In addition, the branch of asymmetric solitons with energy concentrated in the right waveguide channel [the upper dashed branch in Fig. 1(c)] is unstable, but the branch of asymmetric solitons with energy concentrated in the left waveguide channel [the lower dashed branch in Fig. 1(c)] is stable. These stability results are shown in Fig. 2(a), which is a copy of Fig. 1(c) but with stability information displayed (blue for stable and red for unstable). To illustrate the reasons for these stability results, linear-stability eigenvalue spectra for the three solitons in Fig. 1(d,e,f) are plotted in Fig. 2(d,e,f) respectively. These spectra show that the \PT-symmetric soliton in Fig. 1(d), which is above the symmetry-breaking point, is unstable due to the appearance of a pair of real eigenvalues, one of which is positive. The asymmetric soliton in Fig. 1(e), whose energy is concentrated in the right waveguide channel, is unstable due to the appearance of a conjugate pair of eigenvalues on the right half of the complex plane. However, the asymmetric soliton in Fig. 1(f), whose energy is concentrated in the left waveguide channel, is stable because there are no eigenvalues in the right half of the complex plane.

\begin{figure}[htbp]
\centering
\fbox{\includegraphics[width=\linewidth]{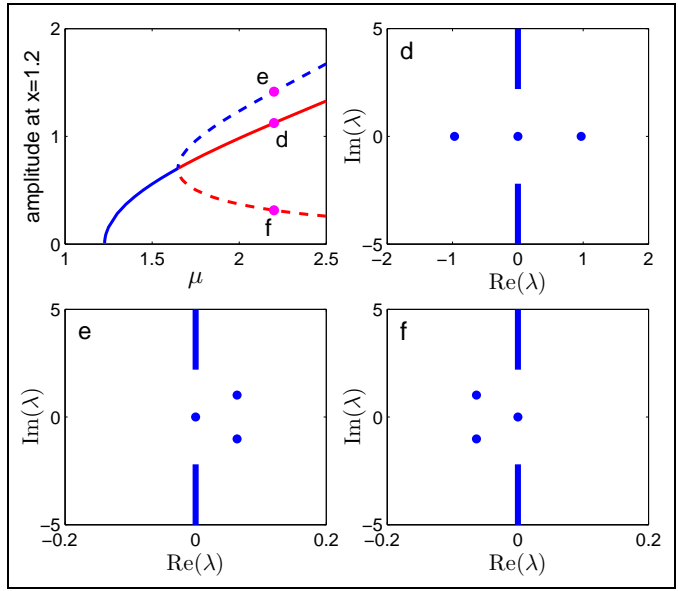}}
\caption{Linear-stability eigenvalue spectra for the three solitons shown in Fig. 1(d,e,f) respectively. The upper left panel is a copy of Fig. 1(c) but with stability illustrated (blue for stable and red unstable).}
\end{figure}

It is constructive to understand the above stability results from the point of view of eigenvalue bifurcations. At the symmetry breaking point $\mu=\mu_c$, zero eigenvalue in the linear-stability operator ${\cal L}$ has geometric multiplicity two and algebraic multiplicity four \cite{Yang2013,YangSAPM}. When the soliton bifurcates away from the symmetry-breaking point, the zero eigenvalue of geometric multiplicity one and algebraic multiplicity two stays at the origin [due to phase invariance of the system (\ref{Eq:NLS})], but the zero eigenvalue with the remaining geometric and algebraic multiplicities bifurcates out of the origin. Along the base branch, since the soliton is \PT-symmetric, if $\lambda$ is an eigenvalue of ${\cal L}$, so must be $\lambda^*, -\lambda$ and $-\lambda^*$. This quartet eigenvalue symmetry forces the zero eigenvalue of algebraic multiplicity two to bifurcate along only the real or imaginary axis, thus creating a pair of real or purely imaginary eigenvalues of opposite sign [see Fig. 2(d)]. Along the bifurcated branches, however, since the soliton is asymmetric, eigenvalues of ${\cal L}$ only feature the pair $(\lambda, \lambda^*)$ symmetry but generally not the quartet symmetry. In this case, that zero eigenvalue of algebraic multiplicity two does not have to bifurcate along the real or imaginary axis. Instead, it generically bifurcates to a conjugate pair of complex eigenvalues on the same side of the complex plane. In particular, when the bifurcation is along the upper branch of Fig. 1(c), the zero eigenvalue bifurcates to the right side, creating oscillatory instability [see Fig. 2(e)]. But when the bifurcation is along the lower branch of Fig. 1(c), the zero eigenvalue bifurcates to the left side, thus not generating any instability [see Fig. 2(f)].

Compared to previous stability results of symmetry-breaking bifurcations of solitons \cite{Panos2005,Kartashov2011,Kirr2011,Yang2013,Yang2014}, the stability switching on the base branch of \PT-symmetric solitons in Fig. 2 is similar to previous results and thus not surprising. However, the opposite stability between the two bifurcated branches of asymmetric solitons in Fig. 2 is totally new because it has never been seen before to the author's best knowledge. Conventional wisdom has always expected the two asymmetric solitons in a symmetry-breaking bifurcation to share the same stability, often out of symmetry considerations of the underlying wave system. However, in \PT-symmetric potentials, that symmetry consideration does not work, which opens the door for opposite stability between asymmetric solitons, as Fig. 2 shows.

To corroborate the linear-stability results in Fig. 2, we consider the nonlinear evolution of these solitons under perturbations. As examples, we perturb the three solitons in Fig. 1(d,e,f) under 10\% random noise initial perturbations, and their subsequent nonlinear evolutions are displayed in Fig. 3(a,b,c) respectively. These evolutions show that the \PT-symmetric soliton above symmetry breaking and the asymmetric soliton with energy concentrated in the right waveguide channel are also nonlinearly unstable, and initial perturbations break them up and drive them to the left waveguide channel. However, the asymmetric soliton with energy concentrated in the left waveguide channel is nonlinearly stable.

\begin{figure}[htbp]
\centering
\fbox{\includegraphics[width=\linewidth]{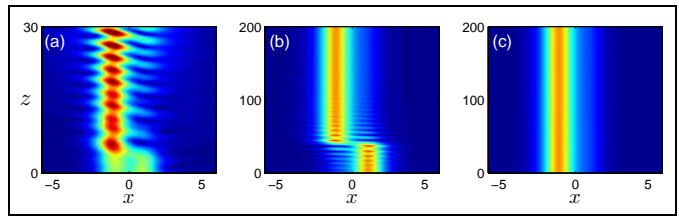}}
\caption{(a,b,c) Nonlinear evolutions of the three solitons in Fig. 1(d,e,f) under 10\% random noise initial perturbations respectively.}
\end{figure}

The opposite stability between the two branches of asymmetric solitons in Fig. 2 and nonlinear evolution results in Fig. 3 reveal that
the wave system (\ref{Eq:NLS}) under the \PT-symmetric potential (\ref{Vx})-(\ref{gx}) favors asymmetric solitons with energy concentrated in the left waveguide channel, and this preference can motivate interesting applications. As an example, one potential application is a unidirectional (non-reciprocal) nonlinear light routing device. In this device, the \PT-symmetric complex potential in Fig. 1(a) is set up (with the refractive index and gain-loss distributions engineered as the real and imaginary parts of this potential), and the waveguide material is chosen to exhibit cubic-quintic nonlinearity as in Eq. (\ref{Eq:NLS}). Then, we launch a Gaussian beam
\[ \label{Gauss}
\Psi(x,0)=R_0 \hspace{0.04cm} e^{-(x-b_0)^2/2}
\]
into this device, where $R_0$ and $b_0$ are the initial amplitude and center position of the beam. At the high amplitude of $R_0=1.2$, evolutions of the beam launched at the initial positions of $b_0=1$ (inside the right waveguide channel) and $b_0=-1$ (inside the left waveguide channel) are displayed in the upper panels of Fig. 4. It is seen that both beams eventually move into the left waveguide channel, exhibiting unidirectional (non-reciprocal) transmission. However, if the initial beam has lower amplitude of $R_0=0.6$, the corresponding evolutions from the above two initial positions are shown in the lower panels of Fig. 4. In this case, transmission is reciprocal and bi-directional. This means that this device exhibits unidirectional transmission only at high amplitudes but not at low amplitudes.

In \cite{Demetri_uni}, unidirectional transmission was reported in a \PT-symmetric dimer (ODE) model. But in that model, unidirectional propagation led to unbounded power growth. In addition, that model only admitted \PT-symmetric stationary modes which do not exhibit symmetry-breaking bifurcations \cite{Li2011}. Thus, the characteristics and mechanisms of unidirectionality between Ref. \cite{Demetri_uni} and this article are very different.

\begin{figure}[htbp]
\centering
\fbox{\includegraphics[width=\linewidth]{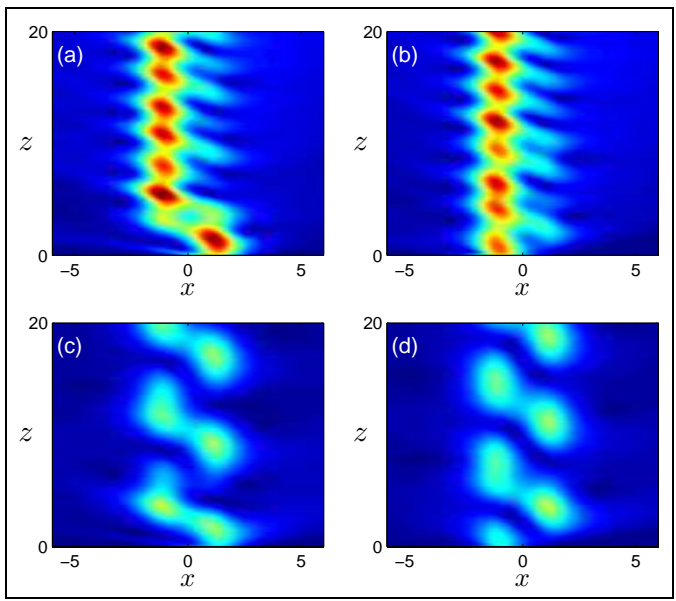}}
\caption{Upper row: unidirectional transmission of high-intensity Gaussian beams (\ref{Gauss}) with $R_0=1.2$ in Eq. (\ref{Eq:NLS}) with the \PT-symmetric potential of Fig. 1(a): (a) launched at $b_0=1$; (b) launched at $b_0=-1$. Lower row: reciprocal transmission of low-power Gaussian beams (\ref{Gauss}) with $R_0=0.6$ in the same model and potential: (c) launched at $b_0=1$; (d) launched at $b_0=-1$.}
\end{figure}

This opposite stability between asymmetric solitons of symmetry-breaking bifurcations in the \PT-symmetric model (\ref{Eq:NLS}) is not restricted to the quintic coefficient $\gamma=-0.1$ used above. We have found that this phenomenon also occurs for a wide range of other positive and negative $\gamma$ values (except for $\gamma=0$ where the nonlinearity is Kerr \cite{Yang2014}). In addition, it occurs for other types of nonlinearities such as saturable nonlinearity. Let us consider the saturable nonlinearity, where Eq. (\ref{Eq:NLS}) becomes
\begin{equation} \label{Eq:NLS2}
\ri \Psi_z + \Psi_{xx} + V(x)\Psi -\frac{E_0}{1+|\Psi|^2}\Psi = 0,
\end{equation}
and $E_0$ is a real constant. Saturable nonlinearity arises in many materials, such as photorefractive crystals \cite{Demetri}.
When the potential $V(x)$ is as given in Eq. (\ref{Vx}), $g(x)$ as given in (\ref{gx}), $\alpha=-0.9$ and $E_0=4$, \PT-symmetry breaking of solitons is displayed in Fig. 5. In this case, we have found that the two branches of asymmetric solitons exhibit opposite stability as well [see Fig. 5(c)]. Thus, this novel type of symmetry breaking can arise in many physical \PT-symmetric systems.

\begin{figure}[htbp]
\centering
\fbox{\includegraphics[width=\linewidth]{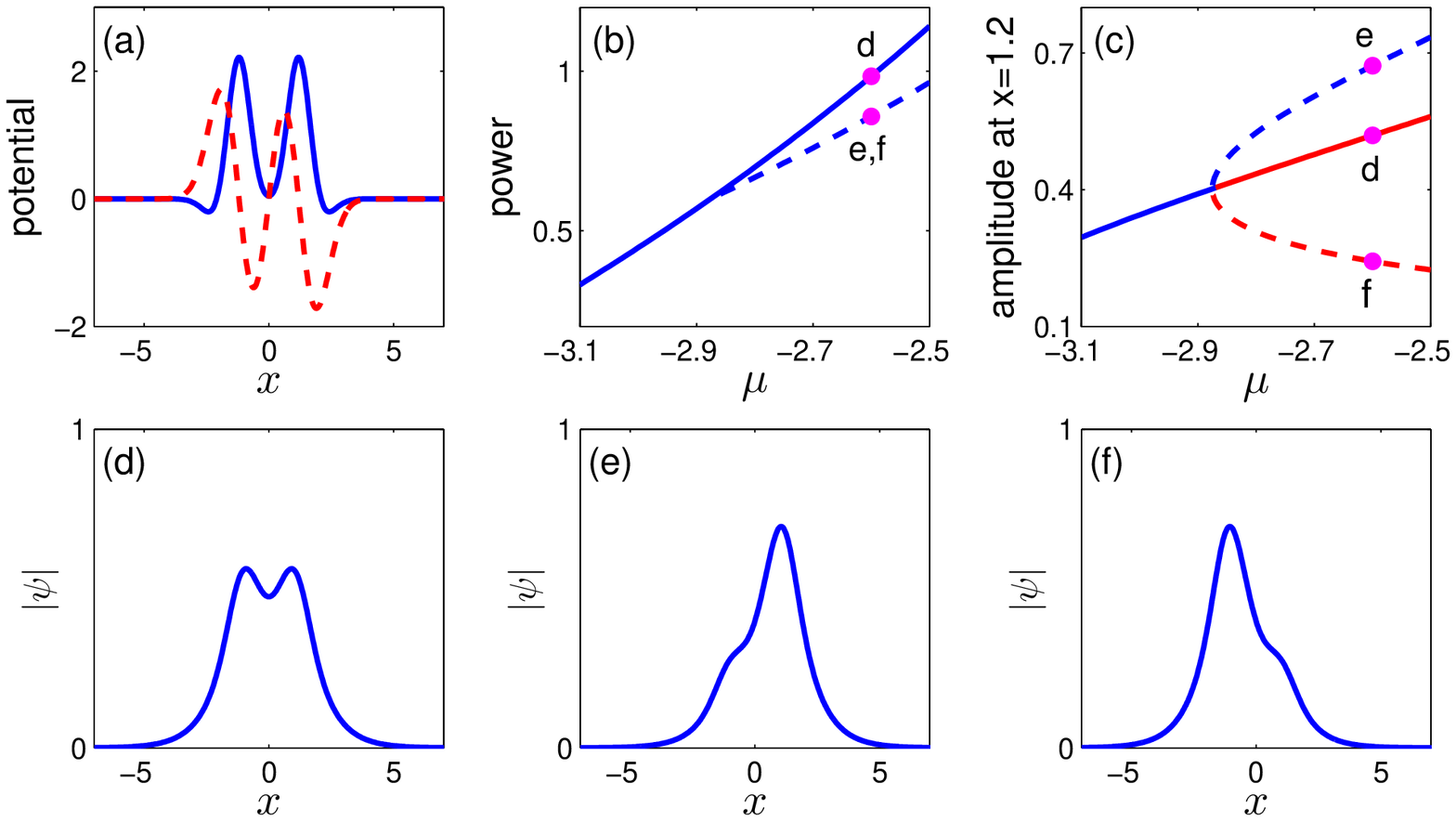}}
\caption{\PT-symmetry-breaking of solitons in the saturable model (\ref{Eq:NLS2}) with $E_0=4$. (a) \PT-symmetric potential $V(x)$ given by equations (\ref{Vx})-(\ref{gx}) with $\alpha=-0.9$; solid blue: Re($V$); dashed red: Im($V$). (b) Power curves of \PT-symmetric (solid) and asymmetric (dashed) solitons in this potential. (c) The amplitude $|\psi|$ of the soliton at the right waveguide channel $x=x_0=1.2$ versus the propagation constant $\mu$. Solid: \PT-symmetric branch; dashed: asymmetric branches. Red color denotes unstable branches, and blue color stable branches. (d,e,f) Profiles of soliton amplitudes $|\psi(x)|$ at parameter points marked by letters `d,e,f' in panels (b,c) respectively.}
\end{figure}

In summary, we have reported a new type of symmetry-breaking bifurcation of solitons in \PT-symmetric optical systems with cubic-quintic and saturable nonlinearities. In this bifurcation, the two bifurcated branches of asymmetric solitons exhibit opposite stability, which contrasts all previous symmetry-breaking bifurcations. This novel bifurcation can be exploited for potential applications, such as devices featuring unidirectional transmission for high-intensity beams but reciprocal transmission for low-intensity beams.

This work was supported in part by the Air Force Office of Scientific Research under award number
FA9550-18-1-0098, and the National Science Foundation under award number DMS-1616122.

\end{document}